\documentstyle[aps,epsfig,multicol]{revtex}
\begin{document}

\newcommand{\onecolm}{
   \end{multicols}
   \vspace{-1.5\baselineskip}
   \vspace{-\parskip}
   \noindent\rule{0.5\textwidth}{0.1ex}\rule{0.1ex}{2ex}\hfill 
	}
\newcommand{\twocolm}{
   \hfill\raisebox{-1.9ex}{\rule{0.1ex}{2ex}}\rule{0.5\textwidth}{0.1ex}
   \vspace{-1.5\baselineskip}
   \vspace{-\parskip}
   \begin{multicols}{2}
	}

\renewcommand{\theequation}{\arabic{equation}}
\def\bra#1{\left\langle{#1}\,\right|\,}    
\def\ket#1{\,\left|\,{#1}\right\rangle}    

\title{Counting statistics for entangled electrons}
\author{Fabio Taddei\thanks{e-mail: Fabio.Taddei@dmfci.unict.it} and Rosario Fazio}
\address{ISI Foundation, Viale Settimio Severo, 65, I-10133  Torino, Italy\\
NEST-INFM \& Dipartimento di Metodologie Fisiche e Chimiche (DMFCI), 
Universit\`a di Catania, Viale A. Doria, 6, I-95125 Catania, Italy}
\date{\today}
\maketitle

\begin{abstract}

The counting statistics (CS) for charges passing through a coherent conductor is the
most general quantity that characterizes electronic transport.
CS not only depends on the transport properties of the conductor
but also depends on the correlations among particles which compose
the incident beam.
In this paper we present general results for the CS of entangled
electron pairs traversing a beam splitter and we show that 
the probability that $Q$ charges have passed is not 
binomial, as in the uncorrelated case, but rather it is symmetric with respect to
the average transferred charge. 
We furthermore consider
the joint probability for transmitted charges of a given spin and we show that
the signature of
entanglement distinctly appears in a correlation which is not present for the non-entangled case.
\end{abstract}

\begin{multicols}{2}

\section{Introduction}
Probably one of the most striking feature of quantum mechanics is 
entanglement~\cite{entanglement} which refers to the nonlocal correlations
existing, even in the absence of interaction, between two (spatially 
separated) parts of a given quantum system. 
Besides the fundamental interest in its generation and detection, a 
great deal of interest has been brought forth by its role 
in quantum information which is attracting a vast effort
due to the very important impact of its potential applications,
ranging from quantum computation to quantum teleportation~\cite{nielsen}.
Entanglement is the main ingredient in all known examples of quantum 
speed-up in quantum computation and communication. 

Most of the work on entanglement has been performed in optical systems 
with photons~\cite{zeilinger}, cavity QED systems~\cite{haroche} and 
ion traps~\cite{wineland}. Only recently people have started to study how 
to generate and to manipulate entangled pairs in a solid state environment.
The prototype setup was discussed in Ref.~\cite{Loss} where it has
been shown  that the presence of spatially separated pairs of entangled 
electrons can be revealed by using a beam splitter, as in Fig. \ref{beam_fig},
and by measuring the correlations of the current fluctuation (noise) at 
the exiting terminals (labeled by 3 and 4 in the figure).
Provided that the  electrons injected into leads 1 and 2 are in an 
entangled state:
\begin{equation}
\ket{\psi}=\frac{1}{\sqrt{2}}\left( \hat{a}^{\dagger}_{2\downarrow}
\hat{a}^{\dagger}_{1\uparrow}
\pm \hat{a}^{\dagger}_{2\uparrow} \hat{a}^{\dagger}_{1\downarrow} \right)
\ket{0} ~,
\end{equation}
bunching and anti-bunching behaviours are found depending on whether state
$\ket{\psi}$ is a spin singlet (lower sign) or a spin triplet (upper sign).
More precisely current noise is enhanced by a factor of 2 with respect to 
non-entangled states in the former case and suppressed to zero in the latter.
Note that while this allows to detect a singlet entangled state, it does
not discriminate between entangled and non-entangled triplets.
Given the general set-up, in order to find  the signatures of entanglement 
in the noise spectrum one needs a physical realization of both the 
{\em entangler} (that enables the pair production) and the beam splitter. 
As the entangler one can resort to 
the phenomenon of Andreev reflection in hybrid normal-superconducting 
systems as discussed in Refs.~\cite{Loss2,Loss3,lesovik}.

Besides electrons, it is possible to produce entangled states with 
Cooper pairs in superconducting nanocircuits~\cite{plastina} or,  
by coupling a mesoscopic Josephson junctions with superconducting 
resonators~\cite{buisson,marquardt}, between Cooper pairs and the 
resonator mode.

In this paper we consider the same approach as in Ref.~\cite{Loss} and
give for granted the existence of an entangler. We address the question 
whether the study of the full statistics of charge transport~\cite{blanter} 
at the exit terminals
3 and 4 of such system can provide
more information (as compared to the noise) on the correlation of the injected 
particles in terminals 1 and 2.
The main result of this paper is that not only the value of the noise
characterizes the entangled singlet state with respect to uncorrelated states
(as shown in~\cite{Loss}), but also the whole probability distribution for 
the transfer of charges is qualitatively modified.
More precisely, we show that the probability distribution relative to
incident particles in the entangled singlet state is not binomial, in contrast
to the case of uncorrelated injected states, and moreover it is symmetric 
around its average value.
In addition, we show that the use of spin-sensitive electron counters, 
on the one hand, provides a more stringent tool for detecting
entangled states which is based on general properties of the
probability distribution.
On the other, it allows to distinguish between entangled and 
non-entangled triplets states.

The paper is organized as follows. In the next section we give a 
brief review of the 
scattering approach for the counting statistics.
Then, in section III, we apply it to the case of the 
beam splitter with entangled electrons. We first present the results for the 
statistics of transmitted charges in a single terminal and then consider the 
cross-correlation. We finally summarize all the results in the Conclusions.

\section{Scattering approach}

In the calculation of counting statistics we adopt the scattering approach of
Landauer and B\"uttiker~\cite{RL,BILP,Bu}. Within this framework, the transport
properties of a metallic phase-coherent structure attached to $n$ reservoirs
are determined by the matrix $S$ of scattering amplitudes.
Such amplitudes are defined through the asymptotic wave functions,
known as scattering states, for particles
in the leads (which connect the reservoirs to the sample).
In one dimension, for example, such scattering states
arising from a unitary flux of particles at energy $E$
originating in the $i$-th reservoir read:
\begin{equation}
\varphi_i(x)=\frac{e^{ik_i(E)x}+r_i(E) e^{-ik_i(E)x}}{\sqrt{hv_i(E)}} ~,
\end{equation}
for the $i$-th lead, and
\begin{equation}
\varphi_j(x)=\frac{t_{ji}(E) e^{-ik_j(E)x}}{\sqrt{hv_j(E)}} ~,
\end{equation}
for the $j$-th lead, with $j\ne i$.
Here $r_i(E)$ is the reflection amplitude
for particles at energy $E$ with wave vector $k_i(E)$ and group
velocity $v_i(E)$ in the $i$-th lead
and $t_{ji}(E)$ is the transmission
amplitude from lead $i$ to lead $j$.
Note that $|r_i|^2$ is the probability for a particle to reflect back into
the $i$-th lead and $|t_{ji}|^2$ is the probability for the transmission
of a particle from lead $i$ to lead $j$.
In the second quantization formalism, the field operator
$\hat{\psi}_{j\sigma}(x,t)$ for spin $\sigma$
particles in lead $j$ is built from scattering states and it is defined as
\begin{equation}
\hat{\psi}_{j\sigma}(x,t)= \int dE ~e^{-\frac{iEt}{\hbar}} \frac{1}{\sqrt{hv_j(E)}}
\left[ \hat{a}_{j\sigma}e^{ik_j x}+\hat{\phi}_{j\sigma}e^{-ik_j x} \right] ~,
\end{equation}
where $\hat{a}_{j\sigma}$ ($\hat{\phi}_{j\sigma}$) is the destruction operator
for incoming (outgoing) particles with spin $\sigma$ in lead $j$.
Such operators are related through the scattering matrix $S$
of the structure as follows:
\begin{equation}
\left( \begin{array}{c} \hat{\phi}_{1\uparrow}\\ \hat{\phi}_{1\downarrow} \\
\hat{\phi}_{2\uparrow} \\ \vdots
\end{array} \right)=~S~
\left( \begin{array}{c} \hat{a}_{1\uparrow}\\ \hat{a}_{1\downarrow} \\
\hat{a}_{2\uparrow} \\ \vdots
\end{array} \right)
\label{Smat}
\end{equation}
and obey anticommutation relations:
\begin{equation}
\left\{ \hat{a}^{\dagger}_{i\sigma}(E),~\hat{a}_{j\sigma'}(E')\right\}
=\delta_{i,j} \delta_{\sigma,\sigma'} \delta(E-E') ~.
\end{equation}
In the case of two and three dimensional leads one can separate 
longitudinal and transverse particle motion. Since the transverse motion 
is quantized, the wave function relative to the plane perpendicular to 
the direction of transport is characterized
by a set of quantum numbers which identifies the channels of the lead. 
Such channels are referred to as open when the corresponding longitudinal 
wave vectors are real,
since they correspond to propagating modes.
Note that the case of a single open channel corresponds to a one dimensional 
lead.

As far as charge transport is concerned, the quantities which are most 
frequently
considered are the conductance and the noise, the latter arising due to the
discrete nature of the charge carriers, even at zero temperature.
However it is more general to consider  the probability
distribution for the transfer of charges~\cite{Lev2,Muz} of which 
conductance and 
noise are the first and second moments, respectively.
Following Refs. \cite{Lev2,Muz}, within the scattering approach
the characteristic function of the probability distribution for the transfer
of particles in a structure attached to $n$ leads at a
given energy $E$ can be written as
\begin{equation}
\chi_E(\vec{\lambda})=\langle \prod_{j=1,n} e^{i\lambda_j(
\hat{N}_I^{j\uparrow} + \hat{N}_I^{j\downarrow})} ~
\prod_{j=1,n}
e^{-i\lambda_j( \hat{N}_O^{j\uparrow} + \hat{N}_O^{j\downarrow})} \rangle ~,
\label{chiE}
\end{equation}
where the brackets $\langle ... \rangle$ stand for the quantum statistical
average in thermal equilibrium.
Assuming a single channel per lead,
$\hat{N}_{I(O)}^{j\sigma}$ is the number
operator for incoming (outgoing) particles with spin $\sigma$ in lead $j$
and $\vec{\lambda}$ is a vector of $n$ real numbers, one for each open channel.
Number operators can be written in terms of the above operators as
$\hat{N}_I^{j\sigma}= 
\hat{a}_{j\sigma}^{\dagger}\hat{a}_{j\sigma}$
and $\hat{N}_O^{j\sigma}=\hat{\phi}_{j\sigma}^{\dagger}\hat{\phi}_{j\sigma}$.
Note that (\ref{chiE}) is simply a generalization of the spinless, single-channel case
for which it is easy to show that
\begin{equation}
\chi_E(\lambda):= \sum_{m,n=0}^1 P_E(m,n)~e^{i\lambda m}~e^{-i\lambda n}=
\langle e^{i\lambda \hat{N}_I}~e^{-i\lambda \hat{N}_O} \rangle ~.
\end{equation}
Here $P_E(m,n)$ is the joint probability for $m$ particles to propagate to the
right and $n$ particles to propagate to the left, with energy $E$.

For long measurement times $t$ \cite{Lev3}, the total characteristic function
$\chi$ is the product of contributions from different energies, so that
\begin{equation}
\chi(\vec{\lambda})=e^{\frac{t}{h}\int
dE~\log{\chi_E(\vec{\lambda})}}
\label{chi}
\end{equation}
and the joint probability distribution for transferring
$Q_1$ electronic charges in lead 1, $Q_2$ in lead 2, etc. is given by:
\begin{equation}
P(Q_1,Q_2,\ldots)=\frac{1}{(2\pi)^n} \int_{-\pi}^{+\pi} d\lambda_1 d\lambda_2
\ldots ~\chi(
\vec{\lambda}) ~e^{i\vec{\lambda}\cdot\vec{Q}} ~.
\end{equation}
In Refs. \cite{Lev1,SS} it was first proved that
in a quantum conductor with a single open channel
the distribution probability
is binomial, in contrast to the
classical case where the distribution is Poissonian.
In Ref. \cite{Yu} the characteristic function was generalized to many open channels
and an explicit expression for its cumulants was obtained.
This allowed to prove that the probability distribution for a tunnel barrier
with very small transmission recovers the Poissonian distribution.
Counting statistics has been so far studied for several systems including hybrid
normal-metal/superconductor structure \cite{Muz,Naz1,Naz2}, metallic diffusive wires
\cite{Yu,Naz} and chaotic cavities \cite{Bl}.

In the rest of the paper we specialize to 
the beam splitter of Fig.\ref{beam_fig}, for which $n=4$.
In analogy with the optical case, we consider the ideal 
situation where particles
injected from branch 1 (2) impinge
on a semi-transparent mirror, from where they are transmitted into 
branch 4 (3) and
reflected into branch 3 (4).

\section{Characteristic function for entangled electrons}

We concentrate in the calculation of  the probability distribution for the
transfer of particles in leads 3 and 4 when particles are injected from leads
1 and 2. Since we are not interested in counting the particles passing
through the entering leads 1 and 2, we set $\lambda_1=\lambda_2=0$, so that Eq.
(\ref{chiE}) becomes
\onecolm
\begin{equation}
\chi_E(\lambda_3,\lambda_4)=\langle e^{i\lambda_3(
\hat{N}_I^{3\uparrow} + \hat{N}_I^{3\downarrow})}~
e^{i\lambda_4( \hat{N}_I^{4\uparrow} + \hat{N}_I^{4\downarrow})}~
e^{-i\lambda_3( \hat{N}_O^{3\uparrow} + \hat{N}_O^{3\downarrow})}~
e^{-i\lambda_4( \hat{N}_O^{4\uparrow} + \hat{N}_O^{4\downarrow})} \rangle ~.
\label{chiE2}
\end{equation}
\twocolm \noindent
We assume, as usual, that the incoming particles are independent
and originate from reservoirs. 
Therefore we set the chemical potentials of reservoirs connected to
leads 3 and 4 to zero and
chemical potentials of reservoirs connected to leads 1 and
2 either to zero or to $eV$.
At zero temperature, the statistical average over the Fermi distribution
function in Eq.(\ref{chiE2}) simplifies to
the expectation value onto the following state
\begin{equation}
\ket{\psi}=\left\{ \begin{array}{c}
\hat{a}^{\dagger}_{1\uparrow} \hat{a}^{\dagger}_{1\downarrow}
\hat{a}^{\dagger}_{2\uparrow} \hat{a}^{\dagger}_{2\downarrow}
\hat{a}^{\dagger}_{3\uparrow} \hat{a}^{\dagger}_{3\downarrow}
\hat{a}^{\dagger}_{4\uparrow} \hat{a}^{\dagger}_{4\downarrow} \ket{0}
~~~\mbox{for $E<0$}\\
\hat{a}^{\dagger}_{1\uparrow} \hat{a}^{\dagger}_{1\downarrow}
\ket{0}
~~~\mbox{for $0<E<eV$}\\
\ket{0}
~~~\mbox{for $E>eV$}
\end{array} \right. ~,
\label{st2}
\end{equation}
in the case where only reservoir 1 is at finite chemical potential $eV$.
The new situation we are interested in corresponds to the propagation of
entangled incident states from branches 1 and 2, as if originating from
an ``entangler''. We describe this device by replacing in (\ref{st2}) the
state
$\hat{a}^{\dagger}_{1\uparrow} \hat{a}^{\dagger}_{1\downarrow} \ket{0}$
with the state
\begin{equation}
\frac{1}{\sqrt{2}}\left( \hat{a}^{\dagger}_{2\downarrow}
\hat{a}^{\dagger}_{1\uparrow}
\pm \hat{a}^{\dagger}_{2\uparrow} \hat{a}^{\dagger}_{1\downarrow} \right)
\ket{0} ~,
\label{stent}
\end{equation}
for $0<E<eV$ and, to ensure no net transfer of charges, we leave unchanged the
state relative to $E<0$ and $E>eV$.
In (\ref{stent}) the minus sign refers to the spin singlet and the plus
sign to the spin triplet.
It is easy to show that for $E<0$ and $E>eV$ one gets
$\chi_E(\lambda_3,\lambda_4)=1$, whereas, for $0<E<eV$, Eq. (\ref{chiE2})
reduces to
\begin{equation}
\chi_E(\lambda_3,\lambda_4)=\langle
e^{-i\lambda_3( \hat{N}_O^{3\uparrow} + \hat{N}_O^{3\downarrow})}~
e^{-i\lambda_4( \hat{N}_O^{4\uparrow} + \hat{N}_O^{4\downarrow})} \rangle ~,
\label{chiE3}
\end{equation}
since states (\ref{st2}) and (\ref{stent}) do not contain incoming particles
from leads 3 and 4.
By using the identity
\begin{equation}
e^{-i\lambda_j \hat{N}_O^{j\sigma}}=\left[ 1+\left(e^{-i\lambda_j}-1\right)
\hat{N}_O^{j\sigma}  \right]
\end{equation}
($(\hat{N}_O^{j\sigma})^2=\hat{N}_O^{j\sigma}$ are projector operators),
the evaluation of $\chi_E(\lambda_3,\lambda_4)$ is reduced to the
calculation of expectation values of number operators and their products.
The procedure is further simplified by assuming no back-scattering
into terminals 1 and 2, so that
the scattering matrix obeys the relation:
\begin{equation}
\left( \begin{array}{c} \hat{\phi}_{3\sigma}\\ \hat{\phi}_{4\sigma}
\end{array} \right)=
\left( \begin{array}{cc}
r_{31}^{\sigma} & t_{32}^{\sigma} \\
t_{41}^{\sigma} & r_{42}^{\sigma}
\end{array} \right)
\left( \begin{array}{c} \hat{a}_{1\sigma}\\ \hat{a}_{2\sigma}
\end{array} \right)
\label{Smat2}
\end{equation}
if  no spin-mixing processes are present.
Here $r_{ij}^{\sigma}$ ($t_{ij}^{\sigma}$) is the reflection (transmission)
amplitude for an incoming particle from lead $j$ to be reflected (transmitted)
into lead $i$.

In the case of entangled incident states (\ref{stent}) we find that
\begin{equation}
\chi_E(\lambda_3,\lambda_4)=\left( \frac{1}{2}-A\right) \left(
e^{-2i\lambda_3} + e^{-2i\lambda_4} \right) + 2A e^{-i(\lambda_3+\lambda_4)}
\label{csent}
\end{equation}
where
\begin{equation}
A=\frac{1}{2} \left[ T^{\uparrow} T^{\downarrow} +R^{\uparrow} R^{\downarrow}
\pm \left( r^\uparrow_{42}
t^{\uparrow\ast}_{41} r^{\downarrow\ast}_{42} t^\downarrow_{41} + 
t^\uparrow_{41}
r^{\uparrow\ast}_{42} t^{\downarrow\ast}_{41} r^\downarrow_{42} \right)
\right] 
\label{A}
\end{equation}
with upper sign referring to the triplet state and
lower sign referring to the singlet state.
$R^{\sigma}=|r_{31}^{\sigma}|^2=|r_{42}^{\sigma}|^2$ and
$T^{\sigma}=|t_{32}^{\sigma}|^2=|t_{41}^{\sigma}|^2$ are reflection and
transmission probabilities, respectively.
Note that the second equalities in the above relationships are completely
general in the case of no back-scattering.
For comparison, in the case of uncorrelated incoming particles described by
the state in Eq.(\ref{st2}) the characteristic function is given by
\begin{equation}
\chi_E(\lambda_3,\lambda_4)=\prod_{\sigma= {\uparrow ,\downarrow}}
	\left( R^{\sigma} e^{-i\lambda_3}+ 
	T^{\sigma} e^{-i\lambda_4} \right) ~.
\label{csord}
\end{equation}

As it appears from Eq.(\ref{csent}) and Eq.(\ref{csord}), the 
characteristic function  relative to entangled pairs of incident 
particles, Eq.(\ref{csent}), possesses a different
structure with respect to the one relative to the ordinary situation 
of independent particles, Eq.(\ref{csord}).
In particular, while Eq.(\ref{csord}) only depends on probability 
coefficients, the characteristic function for entangled electrons
depends directly on the scattering amplitudes.
Furthermore, unlike Eq.(\ref{csent}), Eq.(\ref{csord}) can be 
factorized into spin-up
and spin-down contributions, reflecting the fact that, in the ordinary
situation, electrons with different spin undergo independent 
scattering processes.
In the simplest case of spin-independent transport, such that 
$r_{ij}^{\uparrow}=r_{ij}^{\downarrow}$ and
$t_{ij}^{\uparrow}=t_{ij}^{\downarrow}$, the constant in Eq. (\ref{A}) takes the value
$A=\frac{1}{2}(|t|^2-|r|^2)^2$ for the entangled singlet and
$A=1/2$ for the entangled triplet.
This implies that 
pairs of particles in an entangled triplet state
show the same characteristic function as for non-entangled triplets
(of the form $\ket{\psi}= \hat{a}^{\dagger}_{1\sigma}
\hat{a}^{\dagger}_{2\sigma}$), namely
\begin{equation}
\chi_E(\lambda_3,\lambda_4)=e^{-i(\lambda_3+\lambda_4)} ~.
\label{chi_tr}
\end{equation}
Note, moreover, that the result given in Eq.(\ref{chi_tr}) for 
non-entangled triplets does not depend on transport amplitudes.

It is worthwhile noting that if we allow for spin-polarized transport, 
for example using ferromagnetic metals for terminals 3 and 4, the 
characteristic functions for
all the cases will be distinguished from each other.
The constant $A$ in Eq.(\ref{csent}), in fact, will take the value
\begin{equation}
A=\frac{1}{2}(t^{\uparrow\star}t^{\downarrow}\pm
r^{\uparrow\star}r^{\downarrow}) (t^{\uparrow}t^{\downarrow\star}\pm
r^{\uparrow}r^{\downarrow\star})
\label{Atrip}
\end{equation}
in the case of a symmetric beam splitter
(where $r_{31}^{\sigma}=r_{42}^{\sigma}=r$ and $t_{32}^{\sigma}=t_{41}^{\sigma}=t$).
This makes the characteristic function of the entangled spin triplet 
to differ from the one relative to non-entangled triplets, since in 
the latter case $\chi_E$ is again given by Eq. (\ref{chi_tr}), 
independent of scattering amplitudes.

\subsection{Counting statistics on a single terminal}
Let us now turn the attention to the probability distributions for 
the transfer of particles.
As already mentioned in section II, it can be easily computed by a Fourier
transform of the total characteristic function (\ref{chi}), so that the
probability for transferring a number of $Q_\alpha$ electronic charges,
regardless their spin, into
lead $\alpha$ is given by
\begin{equation}
P(Q_\alpha)=\frac{1}{2\pi} \int_{-\pi}^{+\pi} d\lambda_\alpha ~\chi(
\lambda_\alpha) ~e^{i\lambda_\alpha Q_\alpha} ~.
\end{equation}
Note that $\chi(\lambda_\alpha)$ is obtained from the complete $\chi(\vec{\lambda})$
by setting to zero every $\lambda_\beta$ with $\beta\ne\alpha$.
In the limit of small bias voltage $V$ and zero temperature, the total
characteristic function (\ref{chi})
can be reduced to $\chi(\vec{\lambda})=\left[ \chi_0(\vec{\lambda})\right]^M$
with $M=\frac{eVt}{h}$, in such a way that one only needs to calculate the
characteristic function at zero energy.
In the case of entangled incident particles, state (\ref{stent}),
we find that
\onecolm
\begin{equation}
P(Q_3)=\sum_{k=|Q_3-M|}^M \left( \begin{array}{c} M \\k\end{array} \right)
	\left(\frac{1}{2} -A\right)^k (2A)^{M-k}
	\left( \begin{array}{c} k\\
	\frac{Q_3-M+k}{2} \end{array} \right) ,
\label{pq31}
\end{equation}
\twocolm \noindent
with the sum restricted to values of $k$ such that $(Q_3-M+k)$ is an even
number.
It is easy to show that the distribution (\ref{pq31}) is symmetrical
with respect to the position of the maximum ($Q_3=M$), independently of the
scattering amplitudes.
This result is in contrast with the ordinary situation of independently
injected particles where, as expected, the distribution is binomial:
\begin{equation}
P(Q_3)=\left( \begin{array}{c} 2M \\Q_3\end{array} \right)~R^{Q_3}
\left( 1-R \right)^{2M-Q_3}
\label{21}
\end{equation}
and centered around the value $Q_3=2MR$, for spin-independent transport
(the factor 2 comes from the spin degeneracy).
Note that the width of (\ref{pq31}) for spin singlet is double with respect
to the ordinary case of (\ref{21}) and zero for the triplet.
In particular, for the entangled spin triplet we have
\begin{equation}
P(Q_3)=\delta_{Q_3,M} ~,
\label{pq32}
\end{equation}
equal to the non-entangled triplet states.

Let us now assume spin-dependent transport.
In such a case the distribution $P(Q_3)$
relative to the triplet entangled state broadens to a finite width becoming
distinguished from the two non-entangled triplet states, which remain of the
form (\ref{pq32}).
Such a broadening is due to the fact that the constant $A$ in Eq. (\ref{csent})
is no longer equal to $1/2$, but instead it is given by the expression
(\ref{Atrip}).
As an example, we plot in Fig. \ref{Psp1} the probability
distribution, as a function of the number of charges $Q_3$, relative to the
various incident particle states for a beam
splitter characterized by $R^{\uparrow}=0.2$, $R^{\downarrow}=0.1$ and $M=50$.
The thin (thick) solid line  represents the counting statistics relative to
the entangled singlet (triplet) state, whereas the dashed line is the
counting statistics for the ordinary independent particle state.
The curve relative to the entangled triplet has acquired a
finite width and becomes distinguished from the non-entangled triplets whose
distribution is a Kronecker delta at $Q_3=50$ (not shown in the figure).
Notice that,
since shot noise is proportional to the variance of $P(Q_3)$ through the
relation \cite{blanter} $\frac{s_{33}t}{2e^2}=\ll Q_3Q_3\gg$, where
\begin{equation}
\ll Q_3Q_3\gg =\left. i^2~\frac{\partial^2 \log{\chi(\vec{\lambda})}}{\partial
\lambda_3^2} \right|_{\vec{\lambda}=0} ~,
\end{equation}
we have that
\begin{equation}
s_{33}=\frac{4e^3V}{h} \left( \frac{1}{2}-A\right) ~,
\end{equation}
for entangled particles, and
\begin{equation}
s_{33}=\frac{2e^3V}{h} \left[ R^{\uparrow}\left( 1-R^{\uparrow}\right) 
+R^{\downarrow}\left( 1-R^{\downarrow}\right) \right] ~,
\end{equation}
for independent particles.
For completeness we mention that the dashed curve in Fig. \ref{Psp1},
relative to incoming uncorrelated electrons, corresponds to
the following distribution:
\onecolm
\begin{equation}
P(Q_3)=\sum_{k=\max[0,Q_3-M]}^{\min[Q_3,M]}
\left( \begin{array}{c} M \\k\end{array} \right)~(R^{\uparrow})^{k}
\left( 1-R^{\uparrow} \right)^{M-k}
\left( \begin{array}{c} M \\Q_3-k\end{array} \right)
(R^{\downarrow})^{Q_3-k}
\left( 1-R^{\downarrow} \right)^{M-Q_3+k} ~,
\end{equation}
\twocolm \noindent
which is a convolution of binomial distributions relative to the
two different spin species.

To conclude this section, let us now consider a slightly different situation
in which we suppose to be
able to count the number of electronic charges for
a gives spin, for example by placing a spin-up electron counter on terminal
3 and a spin-down electron counter on terminal 4.
The appropriate expression for the characteristic function reads
\begin{equation}
\chi_E(\lambda_3,\lambda_4)=\langle
e^{-i\lambda_3 \hat{N}_O^{3\uparrow}}~
e^{-i\lambda_4 \hat{N}_O^{4\downarrow}} \rangle ~,
\label{chiE4}
\end{equation}
giving
\onecolm
\begin{equation}
\chi_E(\lambda_3,\lambda_4)=\left( \frac{1}{2} -A \right) \left(
e^{-i\lambda_3} + e^{-i\lambda_4} \right) +A \left[ 1+ e^{-i(\lambda_3+
\lambda_4)} \right]
\label{23}
\end{equation}
\twocolm \noindent
in the case of entangled incident particles from lead 1 and 2.
It is worthwhile noting
that for either $\lambda_3=0$ or $\lambda_4=0$, the function (\ref{23})
is independent of $A$ and, in particular, it is equal for singlet and
triplet states.
This results in the following expression for
the probability of separately counting $Q_3$ spin-up charges
in terminal 3
\begin{equation}
P^{\uparrow} (Q_3)=\left( \begin{array}{c} M\\Q_3 \end{array} \right)
\frac{1}{2^M}
\end{equation}
and $Q_4$ spin-down charges in terminal 4
\begin{equation}
P^{\downarrow} (Q_4)=\left( \begin{array}{c} M\\Q_4 \end{array} \right)
\frac{1}{2^M} ~.
\end{equation}
For completeness, we mention that the characteristic function in the
ordinary case of independent incident particles reads:
\begin{equation}
\chi_E(\lambda_3,\lambda_4)=\left( T^{\uparrow} +R^{\uparrow} e^{-i\lambda_3}
\right) \left( R^{\downarrow} +T^{\downarrow} e^{-i\lambda_4} \right) ~,
\label{chispse}
\end{equation}
which gives the following binomial probability distribution:
\begin{equation}
P^{\uparrow} (Q_3)=\left( \begin{array}{c} M\\Q_3 \end{array} \right)
\left(R^{\uparrow}\right)^{M-Q_3} \left(1-R^{\uparrow}\right)^{Q_3} ~.
\end{equation}
For the non-entangled spin-triplets we have
\begin{equation}
\chi_E(\lambda_3,\lambda_4)= e^{-i\lambda_3} ~,
\end{equation}
which yields
\begin{equation}
P^{\uparrow} (Q_3)=\delta_{Q_3,M} ~.
\end{equation}

\subsection{Counting statistics on both terminals: joint probability}
\label{parc}
Let us now consider
the joint probability for transferring a number of $Q_\alpha$ and $Q_\beta$
electronic charges into, respectively, lead $\alpha$ and $\beta$, given by
\begin{equation}
P(Q_\alpha,Q_\beta)= \int_{-\pi}^{+\pi} 
\frac{d\lambda_\alpha}{2\pi}
\frac{d\lambda_\beta}{2\pi}\chi(
\lambda_\alpha,\lambda_\beta) ~e^{i\lambda_\alpha Q_\alpha +i\lambda_\beta Q_\beta} ~.
\end{equation}
We can distinguish between the two situations: i) spin-insensitive counters with
$\chi_E$ given by (\ref{chiE3}); ii) spin-sensitive counters with $\chi_E$
given by (\ref{chiE4}).
In case i) it is easy to show
that the following relationship holds:
\begin{equation}
P(Q_3,Q_4)=P(Q_3)~ \delta_{2M,Q_3+Q_4}=P(Q_4)~ \delta_{2M,Q_3+Q_4} ~,
\label{2ps}
\end{equation}
which merely expresses the conservation of particles.
Being $2M$ the total number of particles injected from leads 1 and 2 over the
time $t$ and $Q_3$ the number of particles exiting lead 3,
$Q_4=2M-Q_3$ will be the number of particles recorded by counter in 4.
$P(Q_3,Q_4)$, therefore, expresses the correlations due to particles conservation.
This makes explicit the fact that
a measure of the joint probability distribution on terminal 3 and 4 does not
give more information than a
measure of the probability distribution on a single terminal.
The picture changes completely when the constraint of conservation of particles
being counted is lifted, for example, by using spin-selective counters.
This can be realized when a spin-up electron counter is placed on terminal 3 and a
spin-down electron counter on terminal 4.
Note that the number of particles counted is equal to $2M$ only in the case
where there are no spin-down particles exiting lead 3 and no spin-up particles
exiting lead 4.
In the case of pairs of entangled incident particles,
the joint probability of counting $Q_3$ spin-up charges in lead 3
and $Q_4$ spin-down
charges in lead 4 is given by
\onecolm
\begin{equation}
P^{\uparrow\downarrow} (Q_3,Q_4)=\sum_{k=|Q_3-Q_4|}^{\min[Q_3+Q_4,
2M-(Q_3+Q_4)]}\left( \begin{array}{c} M\\k \end{array} \right)
\left(\frac{1}{2} -A \right)^k A^{M-k}
\left( \begin{array}{c} k\\\frac{Q_3-Q_4+k}{2} \end{array} \right)
\left( \begin{array}{c} M-k\\\frac{Q_3+Q_4-k}{2} \end{array} \right) ~,
\label{28}
\end{equation}
\twocolm \noindent
with the sum restricted to values of $k$ such that $[Q_3\pm(Q_4-k)]$ is an even
number.
We see immediately that in the present case Eq.(\ref{2ps}) does not hold and,
in particular, $P^{\uparrow\downarrow} (Q_3,Q_4)$ cannot be expressed in terms
of $P^{\uparrow} (Q_3)$ and $P^{\downarrow} (Q_4)$.
This means, in contrast to case i), that a measure of $P^{\uparrow\downarrow}
(Q_3,Q_4)$ provides more
information than $P^{\uparrow} (Q_3)$ or $P^{\downarrow} (Q_4)$ alone and
reflects the fact that particles counted in terminals 3 and 4 are correlated in a
non-trivial way.
On the contrary, in the ordinary situation of independent
incident particles coming from terminal 1 with $\chi_E$ given by (\ref{chispse})
we have that
\onecolm
\begin{equation}
P^{\uparrow\downarrow} (Q_3,Q_4)=
\left( \begin{array}{c} M\\Q_3 \end{array} \right)
\left(R^{\uparrow}\right)^{M-Q_3}\left(T^{\uparrow}\right)^{Q_3}
\left( \begin{array}{c} M\\Q_4 \end{array} \right)
\left(R^{\downarrow}\right)^{M-Q_4}\left(T^{\downarrow}\right)^{Q_4} ~,
\label{27}
\end{equation}
\twocolm \noindent
which can be written as
\begin{equation}
P^{\uparrow\downarrow} (Q_3,Q_4)=P^{\uparrow} (Q_3) P^{\downarrow} (Q_4) ~.
\end{equation}
Eq. (\ref{27}) confirms that the transfer of spin-up charges into lead 3 and
spin-down charges into lead 4 are independent processes, since the joint
probability is equal to the product of probabilities on individual terminals.
For completeness, we note that when $A=1/2$ in Eq. (\ref{28}), {\it i.e.} the
injected particles are in the entangled triplet states, we have
\begin{equation}
P^{\uparrow\downarrow} (Q_3,Q_4)=
\left( \begin{array}{c} M\\Q_3 \end{array} \right)\frac{1}{2^M}
\delta_{Q_3,Q_4}
\end{equation}
and, when the triplets are non-entangled,
\begin{equation}
P^{\uparrow\downarrow} (Q_3,Q_4)=
\delta_{Q_3,M}\delta_{Q_4,0} ~.
\end{equation}
Remarkably the two expressions above are different even for spin-independent
transport.

The net result is that the relationship between joint probability, on one side, and
single-terminal probabilities, on the other, depends on the specific incident
particle state.
For entangled singlet electrons, in particular, such a relationship does not exist
and furthermore
$P^{\uparrow\downarrow} (Q_3,Q_4)$ depends on the scattering
amplitudes, while $P^{\uparrow} (Q_3)$ does not.
The relevant consequence is that a measure of such a spin-sensitive counting
statistics can provide an unambiguous mean of detecting entangled singlet,
triplet or non-entangled states, since it relies on properties of the
characteristic function rather than on the value of quantities like shot noise.
In practice one should separately measure $P^{\uparrow} (Q_3)$,
$P^{\downarrow} (Q_4)$ and finally $P^{\uparrow\downarrow} (Q_3,Q_4)$ and
compute the ratio
\begin{equation}
r=\frac{P^{\uparrow\downarrow} (Q_3,Q_4)}
{P^{\uparrow} (Q_3)~P^{\downarrow} (Q_4)} ~.
\label{29}
\end{equation}
If $r=1$ independently of $Q_3$ and $Q_4$, we are in the ordinary situation of
independent particles injected either from lead 1 or 2.
If $r=1$ only in the point $(M,0)$ of the $(Q_3,Q_4)$ plane and
zero everywhere else, then we are in the presence
of non-entangled triplets.
If $r\ne 1$, but different from
zero only along the direction $Q_3=Q_4$, we are in the presence of triplet entangled states.
Finally, if
$r\ne 1$ and finite independently of $Q_3$ and $Q_4$ we are in the presence
of a singlet entangled state.
As an example we plot in Figs. \ref{Pq3q4} and \ref{rat} the distribution
(\ref{28}) and
the ratio $r$, respectively, for a singlet entangled state injected in a
spin-independent
beam splitter characterized by $T=0.7$ and $M=50$.
Fig. \ref{Pq3q4} shows that $P^{\uparrow\downarrow}$ possesses an elongated
shape along the direction $Q_4=M-Q_3$, which gets sharper as $T$ goes toward $1/2$.
Fig. \ref{rat} shows that $r$ varies very much in the $(Q_3,Q_4)$ plane:
this allows an easy distinction between different injected particles states.
As a final remark we note that the cross-terminal shot noise in the case of
independent injected particles is zero, whereas in the entangled case is
\begin{equation}
s_{34}^{\uparrow\downarrow}=\frac{2e^3V}{h} \left( A-\frac{1}{4} \right) ~,
\end{equation}
non-zero even for triplets.
This is in contrast with case i) where conservation of counted particles
implies that cross-terminal shot noise is always equal in magnitude
(with opposite sign) to same-terminal shot noise.

\section{Conclusions}
In this paper we have studied the counting statistics
of a beam splitter when
pairs of entangled electrons are injected from the entering terminals 1 and 2.
First we considered the situation in which spin-insensitive electron counters
are placed on terminals 3 and 4.
We found, on the one hand, that the single-terminal probability 
distribution relative to singlet entangled electrons qualitatively 
differs from the one relative to uncorrelated electrons. 
In the former case, in fact, the distribution is not binomial, in contrast 
to the latter case, and furthermore it is symmetric with respect
to the average number of transmitted charges.
On the other hand, we found that the distributions relative to the 
triplet states, both
entangled and non-entangled, are equal and given by unity when the
charge transferred is $M$ and zero otherwise.
Triplet states can be distinguished, however, 
when the transport is spin-polarized, for example when ferromagnetic 
terminals are used.
If this is the case, the single-terminal counting
statistics for the entangled triplet broadens to a finite width, 
while the non-entangled
triplets remains as before.
Interestingly we also noticed that the joint probability for 
counting $Q_3$ electrons arrived
in lead 3 and $Q_4$ electrons arrived in lead 4
does not contain more information than
single-terminal probabilities because of the conservation of particles.
Such a constraint can be lifted by using spin-sensitive electron 
counters, for example
placing a spin-up counter on terminal 3 and a spin-down counter on terminal 4.
In this case the joint probability unambiguously characterizes the state of
the incident electrons.
In particular we found that, unlike in the uncorrelated case, 
in the presence of entanglement the joint  probability cannot be 
expressed as a product of single-terminal
probabilities.
In addition, triplet states exhibit distinguished joint probability depending
on whether they are entangled or not.
Note that the single-terminal counting statistics for the entangled states is also
binomial as for the uncorrelated case, but with probability of the two outcome
equal to $1/2$, therefore independent of scattering amplitudes and total angular
momentum of the pair.
Operatively, we
concluded by showing that the ratio defined in (\ref{29}) can serve as a tool
for discerning among the differently correlated incident electron states.
As shown in paragraph \ref{parc}, a plot of such a ratio as a
function of the number of transferred charges provides an easy and definite way of
identifying entangled singlet, triplet and non-entangled incident states.
We believe that these results can be used for detecting the presence
of entanglement in electronic systems and provide an additional mean for studying
and understanding the production and manipulation of entangled electrons.

\acknowledgments 
The authors would like to thank E. Paladino, G. Falci, G.M. Palma and F. Plastina 
for helpful discussions. This work has been supported by the EU (IST-FET-SQUBIT)
and  by INFM-PRA-SSQI.

\begin{figure}
\begin{center}
\epsfig{figure=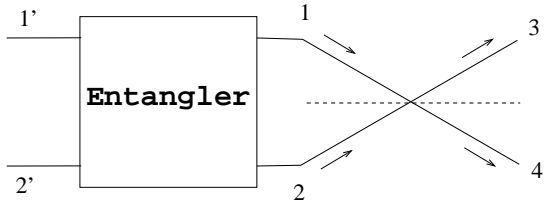,width=0.4 \textwidth}
\end{center}
\caption{The prototype setup consists of an ``entangler'' connected to a beam splitter.
The ``entangler'' produces pairs of entangled electrons from a source of
uncorrelated particles entering from terminals 1' and 2'.
In the beam splitter, the entangled electrons injected in terminals 1 and 2
are transmitted and reflected into terminals 3 and 4 by the
semi-transparent mirror (dashed line). No back-scattering into leads 1 and 2 is
allowed.}
\label{beam_fig}
\end{figure}

\begin{figure}
\begin{center}
\epsfig{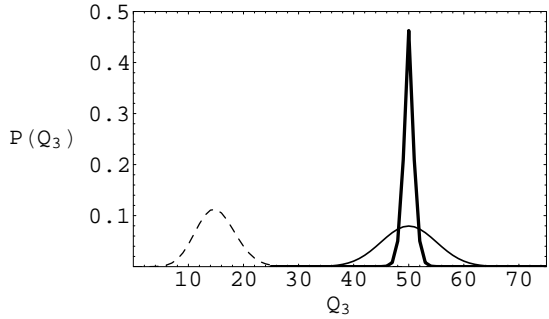}
\end{center}
\caption{Single-terminal counting statistics $P(Q_3)$ for a spin-insensitive
electron counter. Dashed line is relative to uncorrelated electrons, thin line
and bold line are relative to entangled singlet and triplet electrons, respectively. 
The spin-dependent beam splitter is characterized by $R^{\uparrow}=0.2$,
$R^{\downarrow}=0.1$ and $M=50$.}
\label{Psp1}
\end{figure}

\begin{figure}
\begin{center}
\epsfig{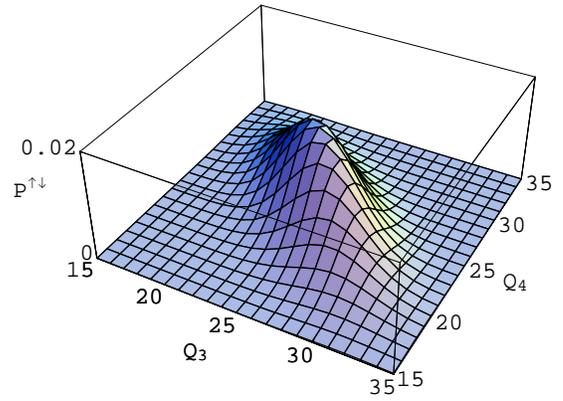}
\end{center}
\caption{Joint probability $P^{\uparrow\downarrow}(Q_3,Q_4)$
for a spin-up electron counter placed on lead 3 and a spin-down
electron counter placed on lead 4. The 3D-plot is relative to
entangled singlet electrons injected from leads 1 and 2.
Beam splitter characterized by $T=0.7$ and $M=50$.}
\label{Pq3q4}
\end{figure}

\begin{figure}
\begin{center}
\epsfig{figure=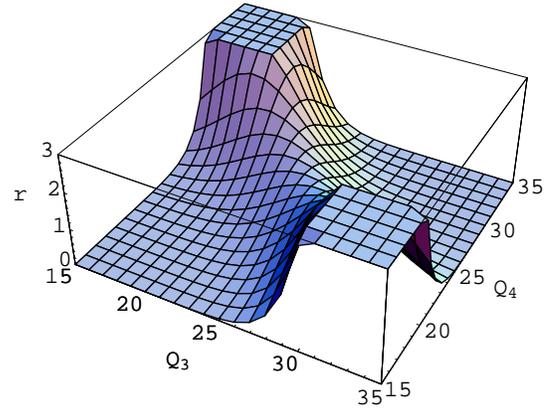,width=0.4 \textwidth}
\end{center}
\caption{3D-plot of the ratio $r(Q_3,Q_4)$ defined in Eq.(\ref{29}) relative
to entangled singlet particles injected in lead 1 and 2. Beam
splitter characterized by $T=0.7$ and $M=50$.}
\label{rat}
\end{figure}

\end{multicols}

\end{document}